%% file: main.tex
\documentclass[conference]{IEEEtran}
\IEEEoverridecommandlockouts
\pdfoutput=1

\usepackage{graphicx}
\usepackage{url}
\usepackage{color}
\usepackage{paralist}
\usepackage{todonotes}
\usepackage{amsmath}
\usepackage{algpseudocode}
\usepackage{caption} 
\usepackage{mdframed}
\usepackage{paralist}
\usepackage{array}

\newcommand{\crit}{\mathit{Crit}}
\newcommand{\replica}{\mathit{Rep}}

\newcommand{\red}[1]{\textcolor{black}{{#1}}}
\begin{document}
  \title{Leveraging SDN to Monitor Critical Infrastructure Networks 
in a Smarter Way\thanks{Work partially supported by EU FP7 project ``Preemptive 
- Preventive Methodologies and Tools to Protect Utilities'', grant no. 607093.}}

  \author{%
    \IEEEauthorblockN{%
      Roberto di Lallo\IEEEauthorrefmark{1},
      Federico Griscioli\IEEEauthorrefmark{1},
      Gabriele Lospoto\IEEEauthorrefmark{1},
      Habib Mostafaei\IEEEauthorrefmark{1},\\
      Maurizio Pizzonia\IEEEauthorrefmark{1} and
      Massimo Rimondini\IEEEauthorrefmark{1}}

    \IEEEauthorblockA{\IEEEauthorrefmark{1}Roma Tre University, Department of 
Engineering\\Via della Vasca Navale 79, 00146 Rome, Italy\\
    \texttt{\{dilallo,griscioli,lospoto,mostafae,pizzonia,rimondini\}@ing.uniroma3.it} 
    }
  }
\maketitle

\begin{abstract}

In critical infrastructures, communication networks are used to exchange vital 
data among elements of Industrial Control Systems (ICSes). Due to the 
criticality of such systems and the increase of the cybersecurity risks in these 
contexts, best practices recommend the adoption of Intrusion Detection Systems 
(IDSes) as monitoring facilities.
The choice of the positions of IDSes 
is crucial to monitor as many streams of data traffic as 
possible. This is especially true for the traffic patterns of ICS networks, mostly confined in many subnetworks, which are geographically distributed and largely autonomous. 
We introduce a methodology and a software architecture that  allow an ICS 
operator to use the spare bandwidth that might be available in over-provisioned networks to forward replicas of traffic streams towards a single IDS placed at an arbitrary location. 
We leverage certain characteristics of ICS networks, like stability of topology and bandwidth needs predictability, and make use of the Software-Defined Networking (SDN) paradigm. We fulfill strict requirements about packet loss, for both functional and security aspects. 
Finally, we evaluate our 
approach on network topologies derived from real networks.

%
\end{abstract}


\input{010-introduction}
\input{015-state-of-the-art}

\input{018-context}

\input{027-requirements}

\input{028-architecture}

\input{030-model}
\input{040-algorithm}

\input{070-topology}

\input{080-variations}
\input{090-conclusions}

\bibliographystyle{IEEEtran}
\bibliography{bibliography}
\end{document}

%% file: 010-introduction.tex
\section{Introduction}
ICSes are the core of critical infrastructures. They are composed by many 
elements that interact by means of a communication network, which we call 
\emph{ICS network}. Main elements of an ICS are embedded devices  that control 
actuators or gather data from sensors. Special servers are in charge to collect 
data from these embedded devices, show them to the control room operators, 
record them in a database, change settings according to operators requests, etc. While the data that flow in an ICS network are very specific, standard 
networking technologies can be adopted for its implementation.

In the past decade, a growth of cyber-attacks directed toward ICSes has been 
observed~\cite{icscertsecattacksreport20092011}. For the security of the ICS 
networks, best practices suggest to deploy network-based 
IDSes~\cite{nistsp80082}. In regular networks it is acceptable to observe 
traffic in a small number of relevant points. However, for reliability reasons, in ICSes, Supervisory Control And Data Acquisition (SCADA) servers are close to sensors and actuators, hence, traffic is mostly local. Further, attacks to ICSes are 
potentially carried out by organizations (e.g., governments, intelligence 
agencies, terrorist groups) that can have insiders and that can carefully design 
attacks so that they pass unobserved by sparsely deployed IDSes. 
Tapping traffic close to all embedded devices and servers can easily lead to 
prohibitive costs. Certain solutions~\cite{nexus7000erspan} make possible to route traffic replicas 
using the 
same ICS network towards one, or a few, IDSes, but they are not able to guarantee the successful delivery of critical ICS  traffic in all cases.

In this paper, we present a methodological approach and an architecture 
to \begin{inparaenum}[(i)]
  \item allow an operator to choose which traffic has to be observed within an ICS network without installing new hardware,
  \item enable the use of the spare bandwidth in the network to forward the traffic to be observed toward an IDS, while avoiding packet loss for regular traffic, and 
  \item guarantee that the IDS receives all the traffic that the operator configured to be observed in order not to introduce false negatives due to packet loss.
\end{inparaenum}
Our solution takes advantage of the fact that topology and bandwidth usage are 
quite stable in ICS networks (see for example~\cite{tom2008recommended}), allowing us to assume in advance knowledge of ICS 
network's traffic, since it derives from ICS design, and to perform a global off-line 
optimization of switching paths. Furthermore, we support the usage of the ICS 
network for additional and occasional traffic, which are always considered 
potentially dangerous. We assume that this traffic can be served with a 
best-effort approach while maximizing the endeavor in observing it. 
We propose an architecture that exploits the Software-Defined 
Network (SDN) approach as prescribed by the OpenFlow 
specifications~\cite{openflowspecification2009version}. 
We evaluated our methodology against four network topologies, derived from real topologies and augmented with realistic 
networks in the domain of
electrical distribution. 
Our experiments show
that our optimization problem can be easily solved for those scenarios in reasonable time and our approach makes efficient use of the bandwidth when the topology allows it.

The rest of the paper is organized as follows. In Section~\ref{sec:soa}, we 
describe the state of the art. In 
Section~\ref{sec:context}, we describe the context of ICSes and introduce basic 
terminologies. In Section~\ref{sec:req}, we formally state the requirements that 
our solution should fulfill. In Section~\ref{sec:architecture}, we describe our 
methodology and our proposed architecture. Section~\ref{sec:problem} introduces 
the ILP formulation for our off-line optimization problem and in 
Section~\ref{sec:algorithm} we show the on-line algorithm for occasional traffic. In 
Section~\ref{sec:evaluation}, we evaluate our approach against realistic 
scenarios. In  
Section~\ref{sec:variations}, we extend our approach in order to relax some simplifying assumptions and handle special cases. Conclusions are drawn in Section~\ref{sec:conclusions and Future Work}.

%% file: 015-state-of-the-art.tex
\section{State of the Art and Background}\label{sec:soa}

ICS networks make use of proprietary protocols, as shown in~\cite{nistsp80082}.
Those protocols (e.g. ModBus~\cite{modbus1996}) are tipically application-layer, 
and they allow the communication among ICS devices. In many 
cases, proprietary protocols are used also to compute routing~\cite{cip}, but this does not limit the adoption of different link-layer technologies~\cite{incibe2015} and new installations tend to be based on 
widely adopted standards, like Ethernet. Protocols adopted in ICS networks do not 
consider security aspects, hence,  well known recommendations 
(e.g.~\cite{nistsp80082}) suggest, among several other countermeasures, the 
adoption of IDSes. 
Forcing network traffic to cross the IDS is not so simple, especially if a 
network administrator needs to be flexible in the selection of traffic that has to be observed. Some flexibility can be gained by adopting 
proprietary protocols (like ERSPAN~\cite{nexus7000erspan}), which however offers 
an unhandy solution and does not guarantee that the rest of the traffic is not affected. 

In the last years, a new centralized approach called 
Software-Defined Networking (SDN) is collecting the attention of the research 
community due to its promising benefits and, in particular, its flexibility in 
the selection of the paths to route packets~\cite{sdnflexible-2013}. 
There have been many attempts in exploiting SDN in security contexts. Some works 
\cite{jw-mdmdusdn-2013,skowyra2013software} propose to implement the IDS as an 
SDN controller module. We argue that such approach poses strong scalability 
issues and it is not advisable in the critical infrastructure context. A 
different approach consists in exploiting SDN to forward traffic towards one or 
more IDSes, as shown in~\cite{shanmugam2014deidtect,jeong2014scalable} for the 
cloud computing applicative context.
These solutions cannot be directly adopted in the ICS context since they do not provide any guarantee about the delivery of regular traffic. 

A relevant aspect in our approach is traffic engineering. In~\cite{roughan2003traffic}, authors show 
that having a traffic-matrix allows traffic engineering problems to be 
easily solved.
Usually, the traffic engineering problem is treated as a \emph{multicommodity 
flow} problem whose solution is described in~\cite{Ahuja:1993:NFT:137406}.  Proposals that are specific to traffic engineering 
for SDN can be found in~\cite{Akyildiz20141,b-ncisrd,agarwal2013traffic}. At the best of our knowledge, our approach is the first attempt to apply traffic engineering to the specific context of traffic monitoring by IDS leveraging the coordinates of the topologies and traffic in ICS networks.

%% file: 018-context.tex
\section{Application Context and Terminologies}\label{sec:context}

For the sake of simplicity, we assume the ICS network to be isolated from the 
corporate network. While this is not completely true in general, still isolation 
(physical or by means of a firewall) is the best practice~\cite{nistsp80082}. 
Hence, in the rest of the paper, we only address traffic monitoring and 
management solely in the context of ICS networks. ICS networks connect several 
kinds of devices. For the purpose of our discussion we divide them in two 
categories. We call the first category \emph{essential}: devices in this 
category can have a very diverse nature, but they are essential for the correct 
operation of the ICS, are part of the ICS design, and are always connected to 
the ICS network. To let the reader better understand the applicative context, we 
provide a more concrete description. We distinguish them in \emph{embedded 
devices} and \emph{servers}. Embedded devices\footnote{For the reader that is 
acquainted with the ICS context, we are referring to Programmable Logic Controllers, 
Remote Terminal Units, Intelligent Electronic Devices, etc.} control 
actuators gather data from sensors, and realize closed-loop control for 
restricted parts of the industrial system. They can send gathered data to 
servers and can be remotely controlled or configured, for example by asking to 
open/close a circuit switcher or by setting values, called \emph{set-points}, 
that are objective of the closed-loop control, like, for instance, a target 
temperature of a heater.
Typically, servers are
\begin{inparaenum}[(i)]
  \item the \emph{SCADA}, which gather data from embedded devices and process them, for example, to detect 
industrial process faults,
  \item the Human-Machine Interfaces (\emph{HMI}) that show to \emph{control 
room operators} the current status of the ICS and allow the operator to specify 
commands or new set-points for embedded devices, and
  \item the \emph{historian} DB, which stores gathered data for future off-line 
analysis.
\end{inparaenum}
We call the second category \emph{non-essential}: occasionally, other devices can be attached to 
the ICS network, for example operators' notebooks to perform maintenance of ICS devices or to perform firmware 
updates.

We call \emph{stream} a communication between two devices on the ICS network. We 
identify it by its source and its destination, specified by IP addresses. Even 
though communications are usually bidirectional, throughout this paper we 
consider a stream to be unidirectional, which means that a full communication 
between two devices generally encompasses two streams. A stream can be critical 
or standard. In a \emph{critical stream}, source and destination are essential
devices and the properties about the stream are known in the ICS design phase. 
In particular their bandwidth demand, source, and destination are known. A 
reliable delivery of critical streams is considered fundamental for the proper 
working of the ICS and substantial resources are available to guarantee this, in 
term of design effort, equipment, etc.  A \emph{standard stream} is not 
essential for the current functioning of the ICS and it is not known in advance. 
It usually involves at least one non-essential device, but it can be involved in 
an occasional communication between two essential devices. Supporting standard 
streams is important to enable occasional use of the ICS network for maintenance 
or other non-critical activities, hence a best-effort delivery is enough for 
this kind of streams.

From the point of view of the security concerns, both kinds of streams are 
equally important, since attacks may involve any of the two with equal chance of 
disruptive effects.
An \emph{attack} to the ICS network consists in any action that introduces 
unexpected traffic or unexpected changes to standard traffic. To be more clear, 
it consists in a source of malicious traffic (e.g. a malware or a rogue device) 
or in the action of tampering with any critical or standard stream. We assume 
that switches cannot be tampered with. We point out that security of switching devices 
is out of the scope of this paper. We suppose there exists a \red{centralized} 
Intrusion Detection System (\emph{IDS}) in the ICS network, which is able to 
recognize malicious traffic and properly send alarms. 

The goal of this paper is to provide a flexible way to use a \red{centralized} 
IDS. To achieve this, we assume that a standard stream $\sigma$ is duplicated, 
generating a \emph{replica stream}; this action is performed at a 
network node that we call \emph{observation point}. Each replica stream $\bar\sigma$, 
associated with $\sigma$, originates at the observation point and ends at the IDS. 
The extension to several IDSes requires minimal effort and it 
is discussed in Section~\ref{sec:variations}.

%% file: 027-requirements.tex
\section{Requirements}\label{sec:req}

In this section, we list the requirements that our methodology should fulfill. We also point out the limitations of the 
current practice. 

\begin{enumerate}
\item\label{req:adaptableOP} \textbf{Observation Points} --  Our methodology
should be able to support the observation of potentially any stream in the network, independently from topology and IDS placement. For security reasons, we prefer observation points close to the destination of streams. 

Concerning current practice, in certain switches, it is possible to remotely mirror a port and also tunneling the traffic of the replica (see for example the ERSPAN technology). However, this approach provides no control on the bandwidth occupation on each link and it is limited to specific vendors support.

\item\label{req:reliableObserved} \textbf{Reliable Replica Forwarding} 
-- Our methodology should guarantee no 
packet loss for replica streams associated with critical or standard streams. 
This is important in order for the IDS to inspect all observed traffic and avoid false negatives due to packet loss.

Concerning current practice, the adoption of remote mirroring technologies implies that the replica is delivered with a best-effort approach. 
To overcome this, in principle, traffic engineering and QoS techniques might be applied. 
However, this considerably increases the architectural complexity. Further, a centralized management, like the one described in Section~\ref{sec:architecture}, is needed anyway.
\item\label{req:reliableCritical} \textbf{Reliable Critical Streams Forwarding} 
--  Our methodology should be able to configure the ICS network so that, for the critical streams, no  packets loss can occur due to congestion.

This requirement is motivated by the fact that, due to Requirement~\ref{req:adaptableOP}, replica streams may easily overload some links and make the usual over-provisioning strategies ineffective.
Actually, up to a certain extent, forwarding reliability can be realized by adopting reliable transport protocols like TCP. However, support of TCP is non-obvious for certain embedded devices. Further, retransmission could introduce a delay that is not acceptable in the ICS context and no bandwidth guarantee is provided. The adoption of QoS and traffic engineering exhibits the same drawbacks as discussed for Requirement~\ref{req:reliableObserved}.

\item\label{req:fairness} \textbf{Standard Streams Usability} -- Our methodology should allow operators to use the ICS network for occasional tasks, which results in injecting new standard streams. 
While the presence of these streams should not adversely impact the fulfillment of other requirements, we expect standard streams to be treated by the ICS network in fair way. Therefore, usage of the ICS network for occasional tasks produce the same outcome for all occasional users and applications.
\end{enumerate}

We also consider the well-founded technology constraint that imposes not to split streams. In fact, if packets of the same stream take different paths, uncontrolled reordering can happen, which is detrimental for TCP performance at best and can change the semantic of datagram-based communications at worst.

%% file: 028-architecture.tex
\section{A Methodology and an Architecture}\label{sec:architecture}

In this section, we describe a methodology and architecture that solve the problem described in Section~\ref{sec:context} with the aim of satisfying the requirements described in Section~\ref{sec:req}.

Our methodology assumes that the network is made of SDN switches that are compliant with the OpenFlow standard~\cite{openflowspecification2009version}. We exploit the OpenFlow features to: 
\begin{inparaenum}[(i)]
\item configure network switches to forward critical streams on the basis of globally optimized paths,
\item configure network switches to forward standard streams on the basis of paths chosen by an on-line greedy approach,
\item instruct certain network switches (\emph{observation points}) to duplicate traffic, for the streams that have to be observed (either critical or standard), and perform the first forwarding step of \emph{replica streams} towards the IDS, 
\item configure network switches to forward replica for critical streams towards the IDS choosing paths that are globally optimized by our off-line approach,
\item configure network switches to forward replicas for standard streams along paths that are dynamically selected with our on-line greedy algorithm, and
\item configure shaping of all streams at ingress network switches. 
\vspace{-5px}
\end{inparaenum}

To meet  Requirements~\ref{req:reliableObserved} and~\ref{req:reliableCritical}, we configure the SDN network to shape each stream at its ingress node, so that packets enter the network at a specified constant rate and all packets exceeding the configured bandwidth are discarded.
For critical streams, the configured maximum bandwidth is determined during the design as described below, so no packet drop should happen. For standard streams, this early limiting avoids congestion of internal nodes that could adversely impact critical streams.
The shaping configuration exploits the \emph{meter} feature of the OpenFlow specifications. 

\begin{figure}
	\centering
	\includegraphics*[width=\columnwidth]{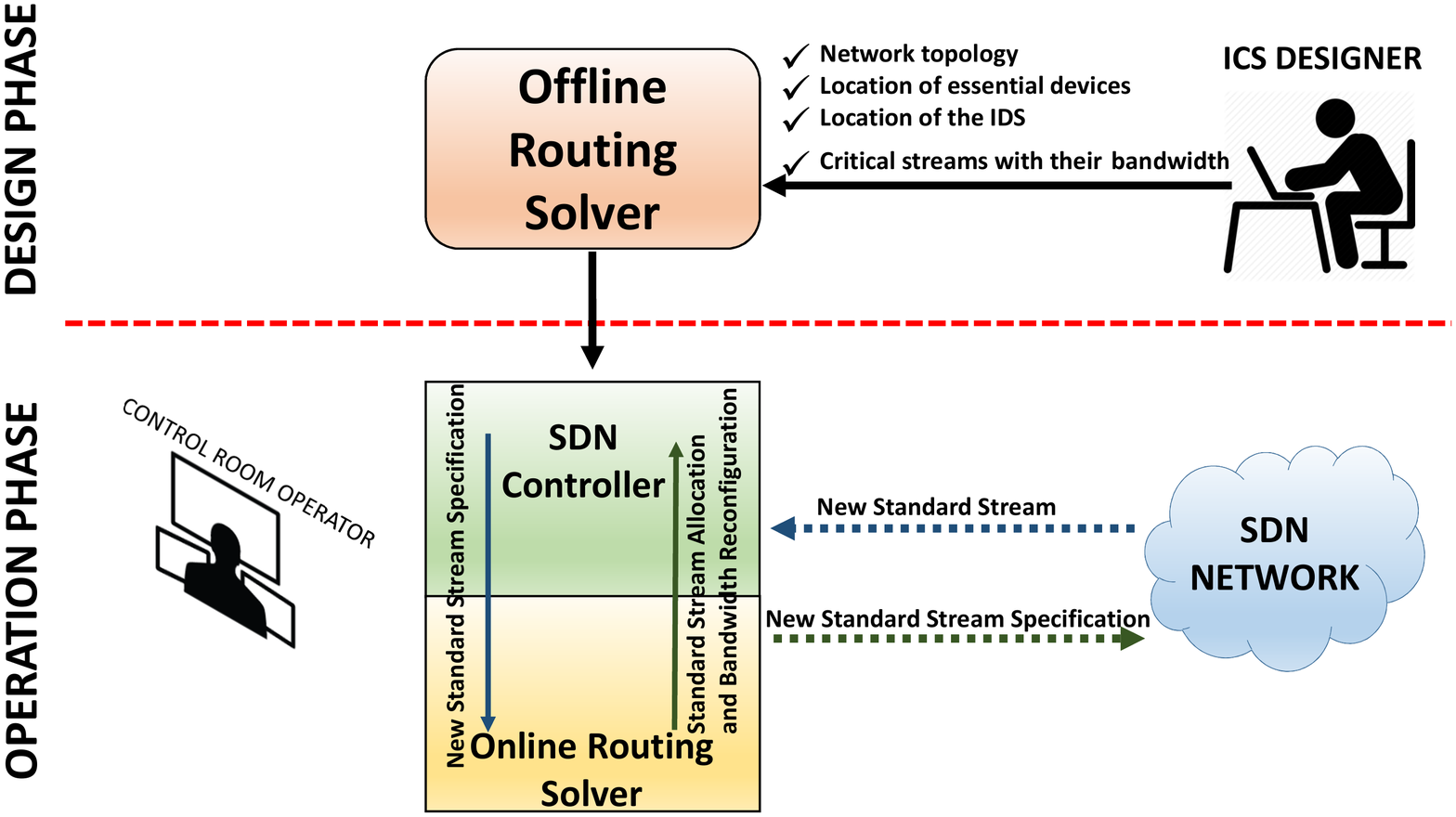}
	    \vspace{-15px}
		\caption{Architecture of our system, with both offline and online routing 
solvers.}
    \vspace{-15px}
	\label{fig:architecture}
\end{figure}

Our methodology encompasses a design phase and an operation phase (see Fig.~\ref{fig:architecture}).
In the \emph{design phase}, we require an \emph{ICS designer} to determine the network topology and to list the critical streams along with their maximal required bandwidth. These data are provided as input to an \emph{off-line routing solver}, which computes the configuration of the SDN switches for critical streams.
More specifically, the input of the off-line routing solver encompasses \begin{inparaenum}[(i)]
\item the network topology,
\item the location of essential devices,
\item the location of the IDS, and
\item for each critical stream its source, its destination and its bandwidth requirement.
\end{inparaenum}
The off-line solver produces, for each critical stream, \begin{inparaenum}[(i)]
\item a forwarding path,
\item an observation point, and
\item a forwarding path for the corresponding replica stream 
starting at the observation point and ending at the IDS.
\end{inparaenum}  The off-line solver is 
based on an ILP formulation, which is described in detail in 
Section~\ref{sec:problem}.

In the \emph{operation phase}, we mandate the adoption of a special architecture (shown in Fig.~\ref{fig:architecture}) in which an \emph{SDN-controller} is in charge of 
configuring forwarding paths and meters to implement shaping. Its configuration is divided into two parts: one for critical streams and one for standard streams. 
The part related to critical streams is configured on the basis of the result of the off-line solver and does not change during operation. 
The part related to standard streams dynamically changes during operation to adapt the configuration of the ICS network when the set of active standard streams changes.
A \emph{control room operator} can monitor the status of the ICS network during production time to have a clear picture of what streams are currently replicated and processed by the IDS.
During operation, any new packet reaching a network switch that does not match any of the rules configured in the switch to forward critical streams
is treated as the first packet of a standard stream $\sigma$.
This packet is forwarded to the SDN-controller as in the classical SDN approach. 
To compute the forwarding path for $\sigma$, 
the SDN-controller takes advantage of an \emph{on-line routing solver}.
This solver shares with the controller the network topology,
and the current available bandwidth on each link derived from currently allocated paths.
It takes as input the source $s$ and destination $t$ of $\sigma$ and computes \begin{inparaenum}[(i)]
\item a forwarding path $P$ for $\sigma$,
\item an observation point $op \in P$ (preferably close to $t$ according to Requirement~\ref{req:adaptableOP}),
\item a forwarding path $Q$ from $op$ to the IDS, and 
\item a new assignment of bandwidth for all standard streams comprising $\sigma$.
\end{inparaenum}   
The details of the on-line routing solver are described in Section~\ref{sec:algorithm}.
These information are used by the controller to re-configure the  shaping for all standard streams but $\sigma$. 
The new standard stream $\sigma$ is configured only after a small amount of time $\tau$ that is dimensioned so that packets related to previous standard streams that where admitted in the network with the old bandwidth allocation are guaranteed to reach destination.

Concerning the path selection, our algorithm has a greedy approach keeping unchanged all paths previously allocated for both kinds of streams.
There are several reasons for this choice:
\begin{inparaenum}[(i)]
	\item sophisticated optimization techniques, like those used in in Section~\ref{sec:problem}, may take a considerable amount of time, which can easily be even larger than the lifespan of the new stream and impair the usability of the network for occasional activities,
	\item modifying the path of a current stream can introduce temporary inconsistencies in the routing that can lead to packet loss, which is against Requirements~\ref{req:reliableCritical} and~\ref{req:reliableObserved},
	\item since standard streams have usually a short lifespan, our main goal is to support them within the requirements listed in  Section~\ref{sec:req}, keeping the  optimization of their resource usage as a secondary goal. 
\end{inparaenum}

%% file: 030-model.tex
\section{Problem Formulation for the Off-Line Routing Solver}\label{sec:problem}

In this section, we present the ILP formulation that is at the basis of the off-line routing solver introduced in Section~\ref{sec:architecture}.
For the sake of simplicity, we made a number of assumptions. Section~\ref{sec:variations} relaxes many of them and describes several extensions.
Our formulation finds,
for each critical stream $\sigma$, a forwarding path $P_\sigma$, an observation point $op_{\sigma}$, and the forwarding path of the replica stream $\bar\sigma$ from $op_{\sigma}$ to the IDS $d$. 
Our formulation is a variation of the well-known multicommodity flow problem~\cite{Ahuja:1993:NFT:137406}. 
In the following, the role of commodities are played by streams and we call \emph{flow} the part of our solution that pertains to a certain critical stream. In this section, all the streams are critical unless different specification is provided. Our variation takes into account the following aspects:
\begin{inparaenum}[(i)]
\item streams are unsplittable, i.e., it is not allowed for a flow to bifurcate (see Section~\ref{sec:req}), 
\item flow demands (i.e., stream bandwidth) are fixed and all critical streams must be routed,
\item each stream can generate a new replica stream originating at its observation point which must be the last traversed node before the destination,
\item nodes of the network that represent embedded devices and servers do not have switching capabilities.
\end{inparaenum}

Since replica streams can take up a lot of bandwidth, we make the observation of a stream optional by introducing a  \emph{relevance} parameter $\rho_\sigma$ for each stream $\sigma$, which indicates how important it is for $\sigma$ to be the observed.

In our formulation, we use the following notation. The network is represented by a  directed graph $G=(V,E)$, where $V$ is a set of vertices and $E$ is a set of directed edges. Each physical link corresponds to two oppositely directed edges $(v,w)$. Each edge $e\in E$ has a capacity $C(e)$ that corresponds to 
the available bandwidth of the link in the corresponding direction.
The set of vertices $V$ is partitioned in two subsets: $N$, representing network switches, and $M$, representing devices with no switching capabilities (e.g., embedded devices and servers). 
We assume that there is no connection among vertices in $M$. The IDS is denoted by $d\in M$. 
For the sake of simplicity,
we do not include the SDN-controller in this model, assuming that connectivity between SDN-controller and network switches is obtained either by a dedicated out-of-band network or by protecting part of the bandwidth of the SDN network using proper configurations.
A stream is a quadruple $\sigma=(s_\sigma,t_\sigma,B_\sigma,\rho_\sigma)$ containing its source, its destination, its bandwidth demand, and its relevance, respectively. A corresponding replica stream is a triple ${\bar\sigma}=(op_\sigma,d,B_{\bar\sigma})$, where 
$op_\sigma$ is its source (such that $(op_\sigma, t_\sigma)\in E$), $d$ is its destination, and 
$B_{\bar\sigma}= B_{\sigma}$ is its bandwidth demand. The set of the critical streams is denoted $\crit$, the set of the corresponding replica streams is denoted $\replica$.

For each $e\in E$ we define $x_\sigma^e \in \{0,1\}$ as a variable that has the following 
meaning 
\begin{equation}\nonumber
   x_\sigma^e =
   \begin{cases}
      1, \quad\textrm{if stream $\sigma$ is being routed through link $e$}\\
      0, \quad\textrm{otherwise}
   \end{cases}
\end{equation}
Analogously, variables $x_{\bar\sigma}^e$ are defined for the corresponding 
replica stream $\bar\sigma$ associated with $\sigma$. 
If a stream $\sigma$ is not observed, it will be $x_{\bar\sigma}^e=0$ $\forall 
e\in E$.

We now define a few convenience functions. We provide definitions for a critical stream $\sigma \in \crit$ and a vertex $v\in V$, the corresponding definitions for replica streams $\bar{\sigma} \in \replica$ are analogous.
\newcommand{\Out}{\textrm{Out}}
\newcommand{\In}{\textrm{In}}
\begin{equation}\label{eq:out}
\mathit{Outgoing\ flow} \qquad \Out_\sigma(v) = \sum_{(v,w)\in E} x_\sigma^{(v,w)}  
\end{equation}
\begin{equation}\label{eq:in}
\mathit{Incoming\ flow} \qquad \In_\sigma(v) =  \sum_{(u,v)\in E} x_\sigma^{(u,v)} 
\end{equation}
\begin{equation}\label{eq:F}
\mathit{Vertex\ flow\  imbalance} \qquad    F_\sigma(v) = \Out_\sigma(v) - \In_\sigma(v)  
\end{equation}
The bandwidth consumed by the critical and replica streams must comply with link capacities: 

\emph{Capacity constraints}.
\begin{equation}\label{eq:capacity}
   \forall e\in E: \quad C(e) - \sum_{\sigma\in\crit} 
(x_\sigma^e+x_{\bar\sigma}^e) \cdot B_\sigma \geq 0
\end{equation}

For each critical or replica streams, we need to express flow conservation. Since flows are unsplittable, each stream generates (consumes) one unit of flow at its source (destination). Conservation is expressed separately for each stream: 

\noindent\emph{Flow conservation and demand constraints for critical streams.}
\begin{equation}
\begin{array}{ll}
 \forall\sigma\in \crit \\
    & \forall v\in V-\{s_\sigma, t_\sigma\}:\quad F_\sigma(v) = 0 \\
    & \Out_\sigma(s_\sigma) = 1, \\
    & \In_\sigma(t_\sigma) = 1
\end{array}
\end{equation}
We now need to express similar constraints for replica streams. 
Let $L_\sigma$ be  the set of the possible observation points for $\sigma$, i.e., 
$L_\sigma=\{v \in N| (v, t_\sigma)\in E \}$. Flows should be balanced for all vertices in $N - L_\sigma$, and each vertex in $L_\sigma$ can produce a unit of replica flow only if it is the last hop of the path assigned to $\sigma$ (by unsplittable flow this is unique), and the IDS cannot be source of flow.

\noindent\emph{Flow conservation and demand constraints for replica streams.}
\begin{equation}\label{eq:replica_conservation_and_demand}
\begin{array}{lll}
 \forall\sigma\in \crit \\
    & \forall v \in N - L_\sigma: \quad F_{\bar\sigma}(v)=0\\
    & \forall v \in L_\sigma:  \quad F_{\bar\sigma}(v) \leq x_{\sigma}^{(v,t)} \\
    & \forall e\in E\ \textrm{exiting}\ d: \quad x_{\bar\sigma}^e=0 \\
\end{array}
\end{equation}
The above constraints also imply that $\In_{\bar{\sigma}}(d) \leq 1$, since for each $\sigma$ only one variable $x_{\sigma}^{(v,t)}$ can be equal to one by the unsplittable flow property. 

As stated above, only vertices in $N$ have switching capabilities. Hence, all nodes in $M$ should have, for their adjacent edges, flow equal to zero but for the streams for which they are source or destination:
\begin{equation}\label{eq:M_does_not_switch_constraint}
\begin{array}{lll}
 \forall\sigma\in \crit, \forall v \in M - \{s_\sigma,t_\sigma\}, e\  \textrm{adjacent to}\  v \\
    \qquad x_\sigma^{e}=0 \\
 \forall\bar\sigma\in \replica, \forall v \in M -\{d\}, e\  \textrm{adjacent to}\  v \\
    \qquad x_{\bar\sigma}^{e}=0 \\
\end{array}
\end{equation}
Our objective function consists of two parts: the first one expresses the 
residual capacity on all the links, while the second states the preference for observing the streams. 
\begin{equation}
\max      \sum_{\sigma\in\crit}\sum_{e \in E}
\frac{C(e)-B_\sigma\cdot(x_{\sigma}^{e}+x_{\bar\sigma}^{e})}{C(e)}  + \sum_{\sigma\in\crit} K\rho_{\sigma} \In_{\bar\sigma}(d)
\end{equation}
Overall, we would like to maximize both parts.
In the above formulation we give precedence to the second part. That is, we 
prefer to observe streams with respect to leaving more residual bandwidth. In order to enforce this, we multiply the second part by 
$K$, which we suppose to be big. We also state that $\rho_{\sigma}$ must be integer and greater than or equal to one, and that $K$ must be chosen to be larger than the range of values that 
the first part can take, namely $K > |E| \cdot |\crit|$.

%% file: 040-algorithm.tex
\section{Standard streams: methodology and algorithm}\label{sec:algorithm}
\newcommand{\bw}{\textrm{bw}}
\newcommand{\none}{\textrm{none}}

\begin{figure}

\textbf{Input:}
\begin{itemize}[-]
\item topology $G(V,E)$ where $V=N\cup M$ (see Section~\ref{sec:problem}),
\item a new  standard stream $\sigma=(s,t)$ with $s,t \in M$,
\item the IDS $d \in N$, 
\item sets $S$ and $C$ of standard and critical streams with \textbf{paths} and 
\textbf{bandwith assignment}.
\end{itemize}
\textbf{Output:}
\begin{itemize}[-]
\item a path $P$ from $t$ to $s$, \item an observation point $op \in P$, 
\item a path $Q$ from $op$ to $d$, 
\item a new bandwidth assignment for streams in $S\cup{\sigma}$.
\end{itemize}
\algrenewcommand{\algorithmicindent}{10pt}
	\small
	\begin{algorithmic}[1]
		
		\ForAll{$e \in E$} \Comment {\footnotesize compute capacities for  $\Call{WidestPath}$ }
			\State Let $m$ be the number of standard streams that shares link $e$ 		
			\State Let $\beta$ be the capacity of $e$ available for standard streams
			\State Assign to each edge $e \in E$ a capacity $C(e)= \beta / (m+1)$ 		
		\EndFor
		\State Let $L(i)$ be the list of vertices in $N$ at distance $i$ from $t$, with $i=1\dots k$, where $k=\textrm{dist}(s,t)-2$.
		
		\State $b_\textrm{best} \gets 0$, $P_\textrm{best} \gets \none$, $Q_\textrm{best} \gets \none$
		
		\For{$i$ in $1\dots k$}
			\ForAll{$v$ in $L(i)$} \Comment $v$ is a candidate observation point
				\State $\textit{SO}\gets \Call{WidestPath}{G, s, v}$
				\State $\textit{OD}\gets \Call{WidestPath}{G, v, d}$
				\State $\textit{OT}\gets \Call{WidestPath}{G, v, t}$
				
				\State $b \gets \min(\bw(OT), \bw(SO), \bw(OD))$
				\If{$b > b_\textrm{best}$}\label{alg:b-bbest-1}
					\State $b_\textrm{best} \gets b$
					\State $op \gets v$
					\State $P_\textrm{best} \gets SO|OT$
					\State $Q_\textrm{best} \gets OD$, 
				\EndIf\label{alg:b-bbest-2}
			\EndFor
			\State \Comment{$op$ is the best observation point at distance $i$ from $t$}
			\If{$b_\textrm{best} > 0$ }\label{alg:wf-1}
				\State Recompute bandwidth assignment for streams	$S\cup{\sigma}$ using Water Filling technique~\cite{radunovic2007unified}.
				\State \Return $P_\textrm{best}$, $op$, $Q_\textrm{best} $,  new 
bandwidth assignment for $S\cup{\sigma}$\label{alg:wf-2}
			\EndIf
		\EndFor	
	\end{algorithmic}
    \caption{Algorithm for handling a new standard stream.}
    \label{fi:on-line_algo}
\end{figure}

In this section, we describe our on-line algorithm for routing standard streams and their related replica streams.
The algorithm takes as input a new standard stream $\sigma=(s,t)$, where $s$ is its source and $t$ is its destination,
and, on the basis of the topology of the network, of the available bandwidth on the links, and of the previously allocated paths and bandwidth, it
produces as result\begin{inparaenum}[(i)]
	\item a path $P$ to be used to forward the packets belonging to $\sigma$,
	\item a switch $op \in P$ (observation point) where the traffic of $\sigma$ is duplicated,
	\item a path $Q$ to be used to forward the replica stream of the traffic of $\sigma$ from $op$ to the IDS,
	\item an assignment of bandwidth for all currently active standard streams, comprising $\sigma$, that should be configured in the ICS network as explained in Section~\ref{sec:architecture}, so that all streams are forwarded respecting Requirements~\ref{req:reliableObserved} and~\ref{req:fairness}.
\end{inparaenum}

Once the path for the new standard stream is computed, our algorithm re-assigns the bandwidth to all standard streams in order to fulfill Requirement~\ref{req:fairness}. 
Bandwidth reduction entails a reconfiguration of limiting and shaping and we assume this operation can be safely performed without any packet loss. However, in order to avoid packet loss during the transition, we should ensure that no queue grows because of the simultaneous presence of packets bursts sent with previous configuration of bandwidth and packets of the new stream $\sigma$, which may account for an overall bandwidth greater than one of the links.

To address this issue, the new stream is admitted in the network only after a small amount of time $\tau$ that ensures that all packets injected with the previous bandwidth configuration are delivered.
The parameter $\tau$ should be greater than the maximum delivery latency of any packet, which, however, is a quite small number and is irrelevant for the vast majority of usage scenarios.
The algorithm is formally described in Figure~\ref{fi:on-line_algo}. As 
motivated in Section~\ref{sec:req}, the algorithm select observation points as 
close as possible to $t$ and secondarily try to allocate the largest possible 
bandwidth. The latter choice takes advantage of the standard 
\textsc{WidestPath()} function~\cite{widest-path1960}, which performs a depth 
first search with backtracking looking for the path with the widest bottleneck. 
Bandwidths to be used in  \textsc{WidestPath()} are computed in the first step 
of the algorithm. To account for bandwidth reassignment for previously allocated 
standard streams, we estimated the bandwidth available for $\sigma$ as the  the 
total bandwidth available for standard streams divided by the number of streams 
after the allocation of $\sigma$.

Then, the algorithm starts enumerating the candidate observation points $op$ 
ordered by increasing distance from $t$. Within the same value of distance, the 
$op$ that allows the widest bandwidth $b$ is chosen. Once $b$ has been 
computed, it is compared with $b_{best}$, replacing it if and only if $b$ is 
greater than $b_{best}$ (lines~\ref{alg:b-bbest-1} --~\ref{alg:b-bbest-2}).
At this point, our algorithm recomputes all bandwidth assignment using the 
Water Filling (WF) technique~\cite{radunovic2007unified} (lines~\ref{alg:wf-1} 
--~\ref{alg:wf-2}), allowing us to find the maximum amount of bandwidth to assign 
to each stream. 
We realize WF in the following way. Suppose, the SDN-controller keeps a data structure that associates with each edge $e$ the set of streams $S(e)$ passing through $e$. Let $c(e)$ be the available bandwidth for standard streams. WF looks for an edge $\bar{e}$ such that  $\bar{e}$ has the minimum of $c(e)/|S(e)|$. WF consider $\bar{e}$ a bottleneck, hence, all streams in $S(\bar e)$ are assigned bandwidth $c(\bar e)/|S(\bar e)|$ and discarded. Remaining bandwidth $c(e)$ are re-computed for all edges and the search is performed again until all streams are discarded and their bandwidth assigned.
In this way, our algorithm successfully computes:
\begin{inparaenum}[i)]
	\item $P_{best}$, namely the best available path;
	\item $op$, namely the starting vertex for replica streams;
	\item $Q_{best}$, namely the best path for replica stream;
	\item new bandwidth assignment for $S$ and $\sigma$. 
\end{inparaenum}

The complexity of the \textsc{WidestPath()} functions is 
$O(|E|)$, as it is based on BFS algorithm, and it is run on each vertex a constant number of times. Hence, the observation point is found in $O(|V||E|)$ time.
The WF takes $O(|E||S|)$.  
Therefore, the overall worst case time complexity of our on-line algorithm is 
$O(|E|(|V|+|S|))$. Actually, in the most common cases, we think the $op$ is found in time much smaller than $O(|V|)$, so the time complexity can be often regarded to be $O(|E||S|)$.

%% file: 070-topology.tex
\section{Evaluation}\label{sec:evaluation}

We validated our approach from three points of view:
\begin{inparaenum}[(i)]
\item we assess the efficiency of our implementation with respect to computation time on realistic instances, inspired by the electricity distribution domain, for both on-line and off-line routing solvers, 
\item we show the efficiency of the bandwidth allocation of the on-line routing solver for standard streams, and
\item we discuss the ability of our solution to meet requirements listed in Section~\ref{sec:req}.
\end{inparaenum}

We identified four different realistic topologies in the following way. We selected four large topologies form topology-zoo.org that are equipped with real link bandwidths or that are fairly mashed. When no links bandwidth are available 1Gbps links was assumed.
We considered each node $n$ to be a router associated with a \emph{city}. We equipped each city with a number of electrical \emph{substations}  whose ICS network is connected to $n$. 
Let $B_n$ be the sum of the bandwidth of all links incident to node $n$. The node with the largest value of $B_n$ is also equipped with one IDS serving the whole network.
\begin{figure}
	\centering
	\includegraphics[width=0.9\columnwidth]{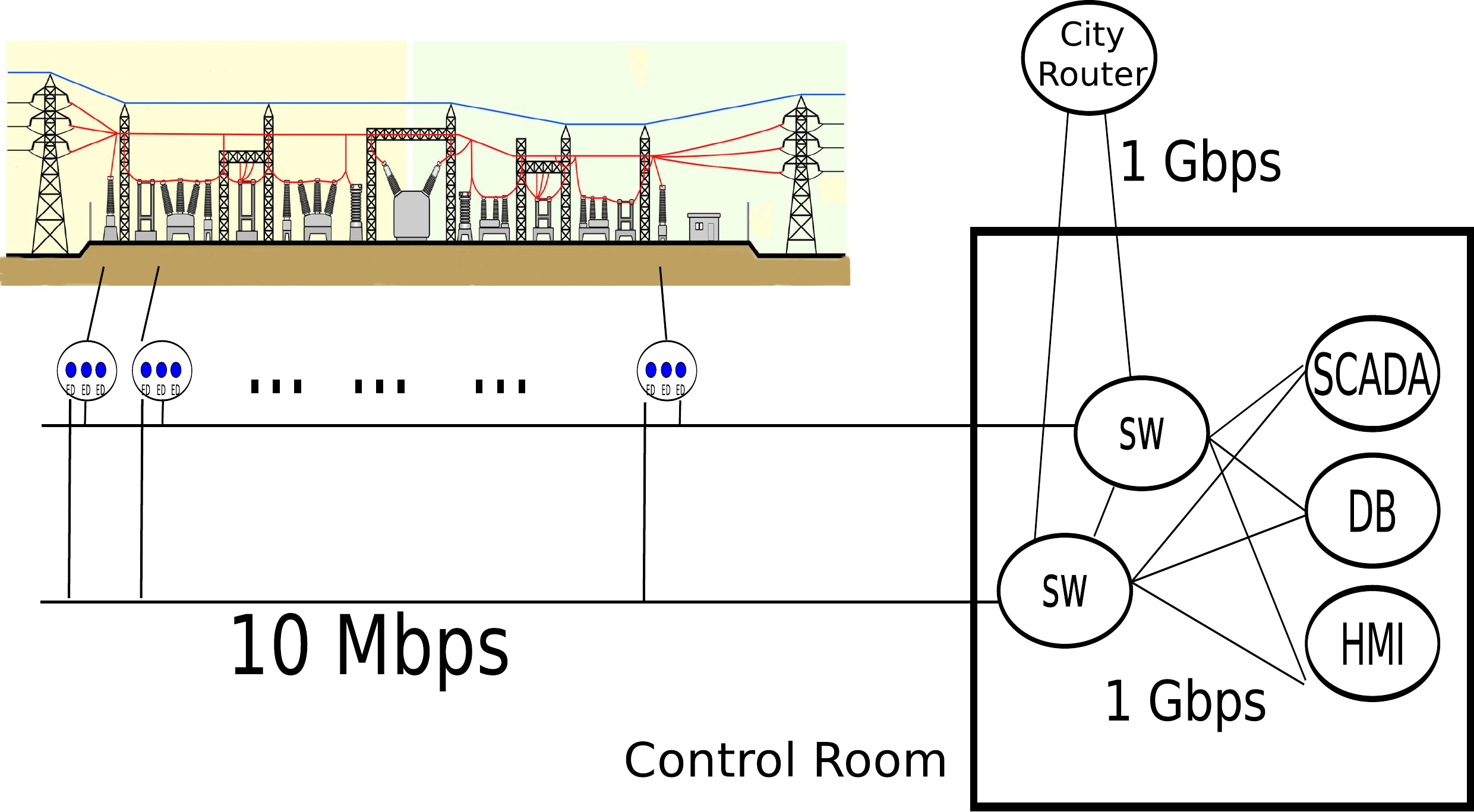}
	\caption{Details of the electricity distribution's substation.}
	\label{fig:evalSubstation}
\end{figure}
The city associated with node $n$, is equipped with $q_n$ identical substations. The total number of substations in the network is $q=\sum_{n} q_n $. The dimensioning of $q_n$ is provided below. The network of a substation is designed on the basis of information that can be freely found in the Internet\footnote{Each of them modeled following the Wikipedia description~\url{https://en.wikipedia.org/wiki/Electrical_substation}}.
Figure~\ref{fig:evalSubstation} shows the topology of a single substation with its connection to the router and Table~\ref{tab:EvaluationTab} shows the devices it contains.
Industrial process data are communicated from embedded devices to the local scada system, and in turn to the HMI and to the DB. The amount of bandwidth required by these communications is shown in Table~\ref{tab:EvaluationTab}, which also show 
the quantity of each  sensors/actuators. For the relevance, we chose always the value 1.
We 
equip each city with a number $q_n$ of substations according to a decreasing power law distribution. In practice, nodes $n$ are sorted by their value of $B_n$. For $n$ with the largest $B_n$, we state $q_n=10$. For $n$ in  position $i$, $q_n=\left\lfloor10/i^\alpha\right\rfloor$, where $\alpha$ is chosen between 0.7 and 1. 
When setting the capacities of the edges we reserved $5\%$ of the bandwidth for standard streams.
Data about used topologies are shown in Table~\ref{tab:topology-data}. 

\begin{table}[]
	\centering
	
	\begin{tabular}{l|c|c|c|}
		
		\cline{2-4}
		&  &   \textbf{From SCADA}& \textbf{To SCADA}   \\ 
		\cline{3-4}
		&\textbf{Qty}  & \textbf{Bandwidth}   &  \textbf{Bandwidth}  \\ \hline
		\multicolumn{1}{|l||}{\textbf{\begin{tabular}[c]{@{}l@{}}Voltage\\ Meter \end{tabular}}}       & 2  & 10 Kbps    & 100 Kbps  \\ \hline
		\multicolumn{1}{|l||}{\textbf{\begin{tabular}[c]{@{}l@{}}Circuit\\ Switches\end{tabular}}}     & 2  & 1.5 Kbps  & 1.5 Kbps \\ \hline
		\multicolumn{1}{|l||}{\textbf{\begin{tabular}[c]{@{}l@{}}Breakers\end{tabular}}}               & 2  & 1.5 Kbps  & 1.5 Kbps \\ \hline
		\multicolumn{1}{|l||}{\textbf{\begin{tabular}[c]{@{}l@{}}Current\\ Meters\end{tabular}}}       & 2  & 10 Kbps  & 100 Kbps \\ \hline
		\multicolumn{1}{|l||}{\textbf{\begin{tabular}[c]{@{}l@{}}Power\\ Transformer\end{tabular}}}    & 1  & 50 Kbps  & 500 Kbps \\ \hline
		\multicolumn{1}{|l||}{\textbf{\begin{tabular}[c]{@{}l@{}}HMI\end{tabular}}}                    & 1  & 30000 Kbps & 3000 Kbps \\ \hline
		\multicolumn{1}{|l||}{\textbf{\begin{tabular}[c]{@{}l@{}}Historian\\ DB\end{tabular}}}         & 1  & 30000 Kbps  & 3000 Kbps \\ \hline

	\end{tabular}
	\caption{Elements of a substation with the bandwidth of the streams used for the evaluation. }
	\label{tab:EvaluationTab}
\end{table}
To validate our off-line routing solver, we instantiated the ILP problem for our four topologies and solved them using Gurobi optimizer ver. 6.5. The formulation set up was performed by using the Python API. The corresponding code is available on the Internet~\cite{sdnci-github}. 
The computation run on a workstation equipped with 8 processors Intel Xeon 2.8GHz.
Results for the off-line solver are shown in Table~\ref{tab:off-line-results}.
The evaluation shows that the formulation of Section~\ref{sec:problem} can be practically used. Considering that the foreseen usage of the formulation is during design, running times are quite small. This makes us believing that our approach could be successfully used even in much larger scenarios.
Even though,  solving times are small, they are not suitable for an on-line use. This justify the introduction of the specific ad-hoc on-line solver, whose algorithm was presented in Section~\ref{sec:algorithm}.

\begin{table}
\centering
{\footnotesize
\begin{tabular}{|c|c|m{0.3cm}|m{0.3cm}|m{0.6cm}|m{0.6cm}|c|p{0.5cm}|c|p{0.6cm}|}
\hline  
   & 
   \multicolumn{5}{|c|}{\textbf{From Topology Zoo}} & 
   \multicolumn{4}{|c|}{\textbf{Input for experiments}} \\
\cline{2-10}  & Name & $|N|$ & $|E|$ &{\tiny 
min bw (bps)} & {\tiny max bw (bps)}&  $q$ & $|N|+|M|$ & $|E|$ & num. strms \\ 
\hline 1 & Cesnet & 10 & 9 & 200M & 600M  & 35  & 501 & 920 & 770 \\
\hline 2 & AttMpls      & 25 &56  & 1G  & 1G  & 50  &726  & 1357  & 1100  \\
\hline 3 & Agis    &25  &30  & 45M &155M  & 42  & 614  & 1123  & 924 \\
\hline 4 & Uninet   & 74 & 101  & 1G  & 1G  & 95  & 1405  & 2572  & 2090 \\

\hline 
\end{tabular} 
}
\caption{Data about original topologies, and topologies used in the experimentation.}
\label{tab:topology-data}
\end{table}

\begin{table}
\centering
{\footnotesize
\begin{tabular}{|c|p{01.2cm}|p{01.2cm}|p{01.2cm}|}
\hline  
   
   \multicolumn{4}{|c|}{\textbf{Results (off-line)}}    \\ 
\hline & gurobi execution time  & number of observed streams & max \%bw on edge\\   
\hline 1 & 12s & 764  &  97.795\%    \\
\hline 2 & 30s & 1100  &   62.060\%  \\
\hline 3 &  33s & 869 & 98.058\%    \\
\hline 4 & 421s & 2087 & 99.455\%   \\

\hline 
\end{tabular} 
}
\caption{Results of the experimentation for the off-line routing solver.}
\label{tab:off-line-results}
\end{table}

To validate the on-line routing solver, for each network, we 
randomly generated a sequence of events (available at~\cite{sdnci-github}) as follows. 
We suppose that standard streams are initiated by (human) operators, whose number is proportional to the network size. We choose to have as many operators as substations (i.e., $q$). Each operator $u$ is attached to a switch $s \in N$ chosen uniformly at random and generates a sequence containing two kinds of events: 
\begin{inparaenum}[(i)]
\item $\mathrm{begin}(c, u, t)$  operator $u$ starts a connection, identified by $c$, with machine $t \in M$, and \item $\mathrm{end}(c)$ connection $c$ ends. \end{inparaenum}
Interarrival time between begin of connections is exponentially distributed with mean $1/\lambda$. Duration of each connection is exponentially distributed with mean $3/\lambda$ (i.e., each operator on average connects to 3 machines at the same time). We set $1/\lambda=5\ \textrm{minutes}$ and the sequence spans about 10 minutes (from 176 to 576 streams).

\begin{figure}
 \includegraphics[width=\columnwidth]{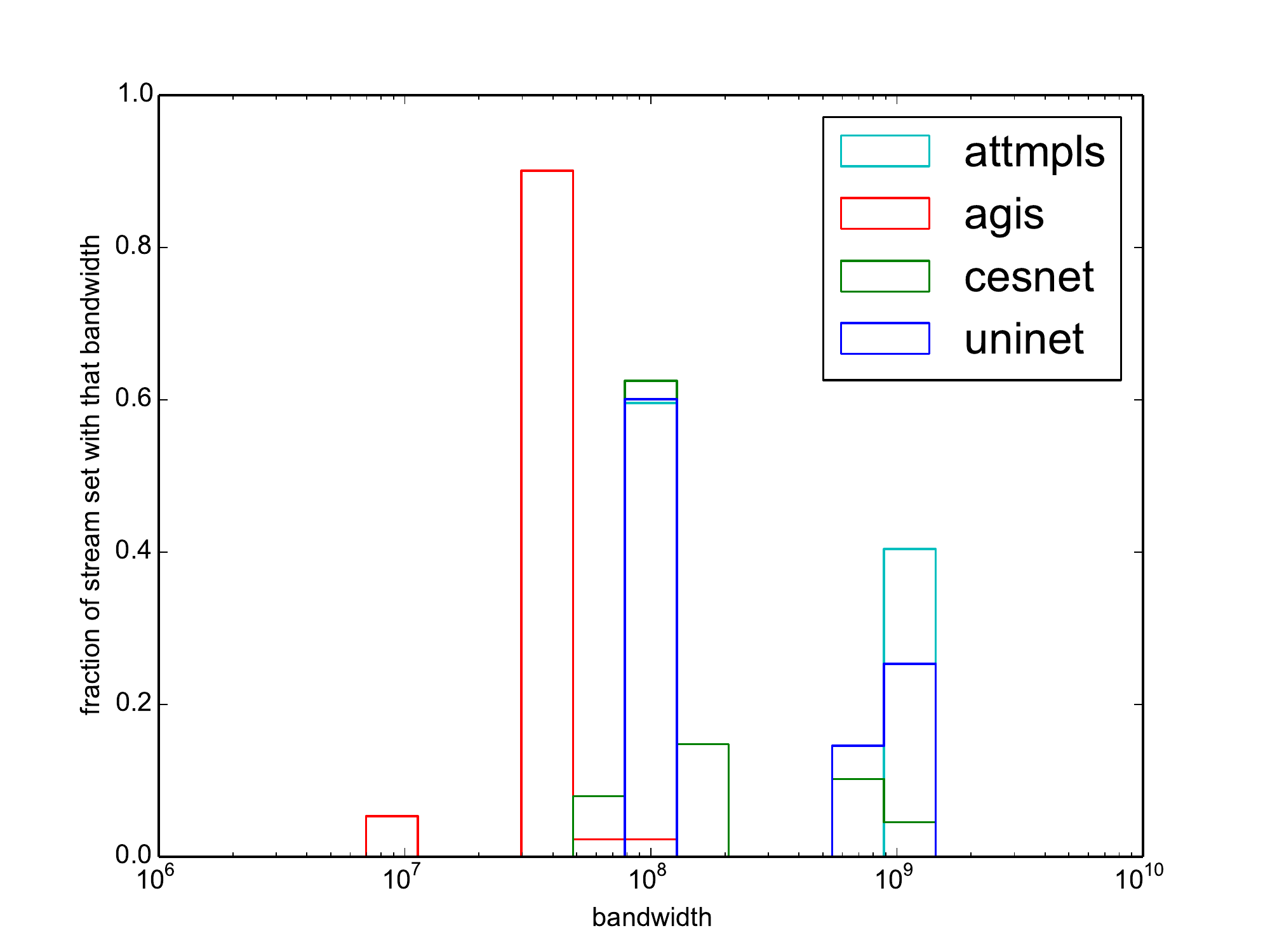}
 \caption{Density of bandwith assigned to streams for each topology (log scale on the x-axis).}
 \label{fig:cdf_online}
 \vskip -0.5cm
\end{figure}

We initialized the status of the solver with the output of the off-line solver for critical streams. Then, we run, for each network, the on-line solver on its sequence of events generated as described above.
Figure~\ref{fig:cdf_online} shows a density diagram, that has on the x-axis possible bandwidth values and on the y-axis the fraction of streams that had that bandwidth assigned in our experiments. In our experiment, assigned bandwidth is always very close to the maximum of the backbone bandwidth. Sometime, if source and destination of the stream are close each other, assigned bandwith can be larger (cf. Table~\ref{tab:topology-data}).

The off-line optimization, together with the traffic shaping approach described in Section~\ref{sec:architecture}, ensures compliance to Requirements~\ref{req:adaptableOP}, \ref{req:reliableObserved}, and~\ref{req:reliableCritical}. 
Further, the inclusion of standard streams is performed only by using the spare bandwidth of each link, thus protecting critical stream and replica streams from packet loss due to congestion (see Section~\ref{sec:algorithm}).
Requirement~\ref{req:fairness} encompasses two essential aspects: fairness of bandwidth allocation  and response time. Our approach handle all streams always assigning the same bandwidth to all of them and dynamically adapting it on the basis of the current needs.
This ensures fairness at expense of some bandwidth waste, since certain streams may not use the whole bandwidth assigned to them. 
To improve this aspect, dynamic polling of bandwidth usage should be adopted~\cite{Akyildiz20141}, however, we believe that in the ICS context, this approach may not be worth the effort.
Concerning response time, this mostly depends on the internal architecture of the SDN-controller. 
A further aspect is the time $\tau$ the controller have to wait to be sure no packet loss occurs when the bandwidth of certain streams have to be reduced (see Section~\ref{sec:architecture}). 
Since $\tau$ should be greater than the time a packet traverse the network, we expect it to be no more than a few milliseconds, which should be negligible for all applications that are reasonable to use in the ICS context.

%% file: 080-variations.tex
\section{Possible Variations and Improvements}\label{sec:variations}

In this section we discuss possible variations to the approach described in 
Sections~\ref{sec:architecture},~\ref{sec:problem}, and~\ref{sec:algorithm}.

\newcommand{\ptitle}[1]{\noindent \textbf{#1.}}

\ptitle{Bandwidth Reservation for Standard Streams}
Our approach statically allocate bandwidth for critical streams and their 
replica streams, using the spare bandwidth for standard streams. However, it is 
easy to use our formulation to explicitly save some bandwidth for this purpose 
during design by artificially reducing the capacities $C(e)$ of 
Constraint~\ref{eq:capacity}.

\ptitle{Dynamicity}
In the description of our approach, we suppose that the needs for monitoring the 
critical streams are known in advance and embodied in the relevance parameters 
$\rho_\sigma$. 
However, there are  situations in which we may want to dynamically 
choose which stream IDS has to analyze.
For example, when an anomaly is recognized, we may want  the IDS analysis to 
focus on the devices close to it, possibly momentarily giving up the inspection 
of traffic of other devices to free up network and IDS resources. This can be 
supported by implementing in the controller with capability to switch off 
observation of critical streams upon request of the control room operator. 
Further, operator may explicitly ask for observation of a critical stream $\sigma$ 
that was currently not observed. To implement this operation,
a search for the widest path starting from the last hop before $t_\sigma$ to the 
IDS have to be performed. If the resulting available bandwidth on the widest 
path is greater than $B_\sigma$, the SDN-controller set up the rules for 
duplication and forwarding toward the IDS, otherwise the search can be done 
backward along the path from the $t_\sigma$ to $s_\sigma$. 
Alternatively, since this somewhat relaxes the support for 
Requirement~\ref{req:adaptableOP}, the bottlenecks identified by the widest path 
algorithm can be used to suggest a set of streams whose observation can be 
switched off to free up enough network resources to satisfy the operator 
request.

\ptitle{Limited IDS Resources}
In our description, we supposed that the IDS has unlimited computational power. 
While this might be reasonable if the IDS is based on cloud technologies, often 
the designer should deal with IDS limits. If we suppose that the IDS is known to 
scale up to a certain bandwidth $B_d$, the formulation of 
Section~\ref{sec:problem} can support it by simply introducing the following 
constraint.
\begin{equation}
\sum_{\bar\sigma\in\replica} \In_{\bar\sigma}(d) \leq B_d
\end{equation}
However, special care should be taken in handling standard streams. In fact, 
during the off-line optimization, some IDS bandwidth should be saved for the 
analysis of standard streams replicas. Further, on-line routing solver must 
consider the IDS bandwidth when calculating the new bandwidth assignment for all 
the standard streams in the WF phase. Essentially, both on-line and off-line 
solver can address the problem as if the IDS were reachable only through a link 
of capacity $B_d$. 

\ptitle{Support for Multiple IDSes}
For the sake of simplicity, in our description, we assumed that only one IDS is 
present in the ICS network. However, there are situations in which it might be 
convenient to have more IDSes $d_1,\dots,d_k \in D$ distributed across the ICS 
network. Hence, a stream can be observed by any of the IDSes. The formulation of Section~\ref{sec:problem} can 
be changed to support this in the following way. Variables $x_{\bar{\sigma}}^e$ are substituted with 
distinct variable sets $x_{\sigma,d}^e$ for each IDS $d\in D$. The functions $\Out_{\sigma,d}(v)$, $\In_{\sigma,d}(v)$, and $F_{\sigma,d}(v)$ are defined for each $d\in D$ as obvious variations of Equations~\ref{eq:out}, \ref{eq:in}, and~\ref{eq:F}. In Constraint~\ref{eq:capacity}, $x_{\bar{\sigma}}^e$ should be substituted by $\sum_{d\in D} x_{\sigma,d}^e$.
Constraints~\ref{eq:replica_conservation_and_demand} should be substituted by
\begin{equation}
\begin{array}{lll}
 \forall\sigma\in \crit \\
    & \forall v \in N - L_\sigma: \quad F_{\sigma,d}(v)=0\\
    & \forall v \in L_\sigma:  \quad \sum_{d\in D} F_{\sigma,d}(v) \leq x_{\sigma}^{(v,t)} \\
    & \forall d\in D, \ \forall e\in E\ \textrm{exiting}\ d: \quad x_{\sigma,d}^e=0 \\
\end{array}
\end{equation}
Since only one variable among $x_{\sigma}^{(v,t)}$ can be greater than zero (by unsplittability of flows), the second inequality implies that only one IDS is involved in the observation of $\sigma$.
The second of Constrants~\ref{eq:M_does_not_switch_constraint} should be substituted by
\begin{equation}
\begin{array}{c}
\forall\sigma\in \crit, \forall d\in D,  \forall v \in M -\{d\}, e\  \textrm{adjacent to}\  v  \\ x_{\sigma,d}^{e}=0 
\end{array}
\end{equation}
Finally, the objective function should be changed into 
\begin{multline}
\max \sum_{\sigma\in\crit} \left(K\rho_{\sigma} \sum_{d\in D}\In_{\bar\sigma}(d) +\right.\\ \sum_{e \in E}
\left.\frac{C(e)-B_\sigma\cdot(x_{\sigma}^{e}+\sum_{d\in D}x_{\sigma,d}^{e})}{C(e)}\right)  
`\end{multline}
With these changes, the formulation automatically perform IDS assignment to streams so that objective function is maximized. 

\ptitle{Flow Table Size Control}
In SDN networks, the number of rules configured in each network switch is a concern. In fact, rules occupy entries in limited size flow tables.
Since, the SDN-controller configures a rule for each outgoing stream, limits to the flow table can be take into account by the following constraints, where $FT(v)$ is the maximum number of rules that can be configured in the switch $v$.
\begin{equation}
\forall v \in N  \sum_{\sigma\in \crit} \left( \Out_{\sigma}(v)+ \sum_{\forall d\in D} \Out_{\sigma,d}(v) \right) \leq FT(v)
\end{equation}

%% file: 090-conclusions.tex
\section{Conclusions}\label{sec:conclusions and Future Work}

We proposed a methodology and an architecture that enable flexible adoption of one IDS (or a few of them), while keeping the possibility to mirror any stream in the network and forward it toward the IDS independently from its deployment location. While we think that our approach can be useful in many contexts, we tailored it for the usage within ICS networks, where most of the traffic flows are critical and known in advance, and occasional usage can be handled with a best effort approach. We base our work on SDN technology, which allowed us to keep a simple centrally managed network configuration.
We presented several small-effort extensions to the basic description in Section~\ref{sec:variations}. However, the integration of a distributed approach for the SDN-controller, like the one presented in~\cite{bib:hyperflow-2010}, in our architecture, may be the subject of additional research. 
Further, in our solution, we statically assigned bandwidth to all critical streams, disregarding cases in which traffic is not stable over time. Better usage of the bandwidth could be achieved by taking this into account.

%% file: main.bbl
\begin{thebibliography}{10}
\providecommand{\url}[1]{#1}
\csname url@samestyle\endcsname
\providecommand{\newblock}{\relax}
\providecommand{\bibinfo}[2]{#2}
\providecommand{\BIBentrySTDinterwordspacing}{\spaceskip=0pt\relax}
\providecommand{\BIBentryALTinterwordstretchfactor}{4}
\providecommand{\BIBentryALTinterwordspacing}{\spaceskip=\fontdimen2\font plus
\BIBentryALTinterwordstretchfactor\fontdimen3\font minus
  \fontdimen4\font\relax}
\providecommand{\BIBforeignlanguage}[2]{{%
\expandafter\ifx\csname l@#1\endcsname\relax
\typeout{** WARNING: IEEEtran.bst: No hyphenation pattern has been}%
\typeout{** loaded for the language `#1'. Using the pattern for}%
\typeout{** the default language instead.}%
\else
\language=\csname l@#1\endcsname
\fi
#2}}
\providecommand{\BIBdecl}{\relax}
\BIBdecl

\bibitem{icscertsecattacksreport20092011}
I.~control systems cyber emergency response team control systems~security
  program, ``Ics-cert incident response summary report 2009-2011,'' ICS-CERT,
  Tech. Rep., 2011.

\bibitem{nistsp80082}
K.~Stouffer, S.~Lightman, V.~Pillitteri, M.~Abrams, and A.~Hahn, ``Guide to
  industrial control systems (ics) security -- nist special publication (sp)
  800-82 revision 2,'' NIST, Tech. Rep., 2015.

\bibitem{nexus7000erspan}
``Cisco nexus 7000 series nx-os system management configuration guide,'' Cisco
  Systems Inc., Tech. Rep., 2011.

\bibitem{tom2008recommended}
S.~Tom, D.~Christiansen, and D.~Berrett, ``Recommended practice for patch
  management of control systems,'' \emph{DHS control system security program
  (CSSP) Recommended Practice}, 2008.

\bibitem{openflowspecification2009version}
O.~S. Specification, ``Version 1.3.3 (wire protocol 0x04),'' Sept 2013.

\bibitem{modbus1996}
I.~A.~S. MODICON, Inc., ``Modbus protocol -- reference guide,'' Tech. Rep.,
  1996.

\bibitem{cip}
I.~ODVA, ``The common industrial protocol (cip),''
  https://www.odva.org/Technology-Standards/Common-Industrial-Protocol-CIP/.

\bibitem{incibe2015}
M.~Herrero~Collantes and A.~López~Padilla, ``Protocols and network security in
  ics infrustructures,'' INCIBE, Tech. Rep., 2015.

\bibitem{sdnflexible-2013}
H.~Kim and N.~Feamster, ``Improving network management with software defined
  networking,'' \emph{IEEE Communications Magazine}, vol.~51, no.~2, pp.
  114--119, February 2013.

\bibitem{jw-mdmdusdn-2013}
\BIBentryALTinterwordspacing
R.~Jin and B.~Wang, ``Malware detection for mobile devices using
  software-defined networking,'' in \emph{Proceedings of the 2013 Second GENI
  Research and Educational Experiment Workshop}, ser. GREE '13.\hskip 1em plus
  0.5em minus 0.4em\relax Washington, DC, USA: IEEE Computer Society, 2013, pp.
  81--88. [Online]. Available: \url{http://dx.doi.org/10.1109/GREE.2013.24}
\BIBentrySTDinterwordspacing

\bibitem{skowyra2013software}
R.~Skowyra, S.~Bahargam, and A.~Bestavros, ``Software-defined ids for securing
  embedded mobile devices,'' in \emph{High Performance Extreme Computing
  Conference (HPEC), 2013 IEEE}.\hskip 1em plus 0.5em minus 0.4em\relax IEEE,
  2013, pp. 1--7.

\bibitem{shanmugam2014deidtect}
P.~K. Shanmugam, N.~D. Subramanyam, J.~Breen, C.~Roach, and J.~Van~der Merwe,
  ``Deidtect: towards distributed elastic intrusion detection,'' in
  \emph{Proceedings of the 2014 ACM SIGCOMM workshop on Distributed cloud
  computing}.\hskip 1em plus 0.5em minus 0.4em\relax ACM, 2014, pp. 17--24.

\bibitem{jeong2014scalable}
C.~Jeong, T.~Ha, J.~Narantuya, H.~Lim, and J.~Kim, ``Scalable network intrusion
  detection on virtual sdn environment,'' in \emph{Cloud Networking (CloudNet),
  2014 IEEE 3rd International Conference on}.\hskip 1em plus 0.5em minus
  0.4em\relax IEEE, 2014, pp. 264--265.

\bibitem{roughan2003traffic}
M.~Roughan, M.~Thorup, and Y.~Zhang, ``Traffic engineering with estimated
  traffic matrices,'' in \emph{Proceedings of the 3rd ACM SIGCOMM conference on
  Internet measurement}.\hskip 1em plus 0.5em minus 0.4em\relax ACM, 2003, pp.
  248--258.

\bibitem{Ahuja:1993:NFT:137406}
R.~K. Ahuja, T.~L. Magnanti, and J.~B. Orlin, \emph{Network Flows: Theory,
  Algorithms, and Applications}.\hskip 1em plus 0.5em minus 0.4em\relax Upper
  Saddle River, NJ, USA: Prentice-Hall, Inc., 1993.

\bibitem{Akyildiz20141}
\BIBentryALTinterwordspacing
I.~F. Akyildiz, A.~Lee, P.~Wang, M.~Luo, and W.~Chou, ``A roadmap for traffic
  engineering in sdn-openflow networks,'' \emph{Computer Networks}, vol.~71,
  pp. 1 -- 30, 2014. [Online]. Available:
  \url{http://www.sciencedirect.com/science/article/pii/S1389128614002254}
\BIBentrySTDinterwordspacing

\bibitem{b-ncisrd}
G.~B\'ela, ``{Networked Critical Infrastructures}: Secure and resilient by
  design,'' in \emph{The Proceedings of the EUROPEAN INTEGRATION BETWEEN
  TRADITION AND MODERNITY Congress}, vol.~6, 2015, pp. 753--760.

\bibitem{agarwal2013traffic}
S.~Agarwal, M.~Kodialam, and T.~Lakshman, ``Traffic engineering in software
  defined networks,'' in \emph{INFOCOM, 2013 Proceedings IEEE}.\hskip 1em plus
  0.5em minus 0.4em\relax IEEE, 2013, pp. 2211--2219.

\bibitem{radunovic2007unified}
B.~Radunovi{\'c} and J.-Y.~L. Boudec, ``A unified framework for max-min and
  min-max fairness with applications,'' \emph{IEEE/ACM Transactions on
  Networking (TON)}, vol.~15, no.~5, pp. 1073--1083, 2007.

\bibitem{widest-path1960}
M.~Pollack, ``Letter to the editor—the maximum capacity through a network,''
  \emph{Operations Research}, vol.~8, no.~5, pp. 733--736, 1960.

\bibitem{sdnci-github}
``{\textbf{Companion Website with code}},''
  \\\url{https://bitbucket.org/sdnci/sdn-ci/}.

\bibitem{bib:hyperflow-2010}
\BIBentryALTinterwordspacing
A.~Tootoonchian and Y.~Ganjali, ``Hyperflow: a distributed control plane for
  openflow,'' in \emph{Proceedings of the 2010 internet network management
  conference on Research on enterprise networking}, ser. INM/WREN'10.\hskip 1em
  plus 0.5em minus 0.4em\relax Berkeley, CA, USA: USENIX Association, 2010, pp.
  3--3. [Online]. Available:
  \url{http://dl.acm.org/citation.cfm?id=1863133.1863136}
\BIBentrySTDinterwordspacing

\end{thebibliography}
